\documentclass[a4paper]{article}
\usepackage{amsmath, amsfonts}

\usepackage{amssymb}
\usepackage{enumerate}

\providecommand{\keywords}[1]
{
  \small	
  \textbf{\textit{Keywords:}} #1
}

\newtheorem{theorem}{Theorem}

\newtheorem{corollary}[theorem]{Corollary}

\newtheorem{lemma}{Lemma}

\title{Regular Dynamics and Collisions Inside Classical Closed String}
\author{V.A.~Malyshev$^{1}$, A.A.~Zamyatin$^{1}$  \\
\small 
\textit{$^{1}$Lomonosov Moscow State University} \\
}
\date{}

\begin{document}
\maketitle


\begin{abstract}  
We consider classical closed string with $N$ particles inside.
Taking collisions into account, we consider dynamics of this $N$-particle
system under the influence of constant external force. We
get Euler equations and explicit formula for the pressure.
\end{abstract}

\tableofcontents

\section{Introduction}

In 1950--1990 there was the great explosion of mathematical activity
in equilibrium statistical physics. Its mathematical success was mainly
due to the only axiom (Gibbs distribution for many particle systems)
and quite new and beautiful mathematical problems. But now this activity
slows down. It can only mean that transition to non-equilibrium mathematical
statistical physics is necessary. It seems to be much more difficult
-- no evident axioms, much less probability theory. And dynamics in
multi-particle systems is more difficult to rigorously analyze. Its
starting period can consist of the following steps:

1. Convergence (as $t\to\infty$) to equilibrium for systems with
large but fixed number $N$ of particles with purely deterministic
interaction and minimum stochasticity in external influence, see [1--6].

2. Convergence (as $t\to\infty$) to stationary flows where $N$ particles
move under deterministic or random external forces, see [7,8].

3. Convergence (as $N\to\infty$) of $N$ particle systems to regular
continuum particle systems and rigorous deduction of Euler equations
for these limiting continuum particle systems, see [9,10].

Here we consider problems 2 and 3 for flows of particles on the circle. 

\subsection{The Model}

Consider the set $X_{\infty}=X_{\infty}(N.L)$ of infinite periodic
sequences of point particle coordinates on the real axis $R$ 
\begin{equation}
\ldots <x_{-1}<x_{0}<x_{1}<\ldots <x_{N-1}<x_{N}<\ldots ,\label{infinite}
\end{equation}
where periodic means that $x_{k+N}=x_{k}+L$ for any $k$, some fixed
real $L>0$ and integer $N>0$. We assume that locally the dynamics
is defined by Newton's equations (masses are assumed to be $1$ as
mass scaling can be absorbed by scaling of other parameters $\omega,\alpha,f_{k}$)
\begin{equation}
\ddot{x}_{k}=\omega^{2}(x_{k+1}-2x_{k}+x_{k-1})-\alpha\dot{x}_{k}+f_{k},\label{x_equations}
\end{equation}
with external (driving) forces $f_{k}=f_{k}(t)=f_{k+N}(t)$, dissipative
forces $-\alpha\dot{x}_{k},\alpha\geq0,$ and formal interaction potential
energy 
\begin{equation}
U_{a}=\frac{\omega^{2}}{2}\sum(x_{k+1}-x_{k}-a)^{2},\label{potential}
\end{equation}
that formally gives the same equations for any $a$. 

In this paper we consider only the case when for all $k$
\begin{equation}
f_{k}(t)=f(t)\label{f_t_f}
\end{equation}
for some function $f(t)$. Unless otherwise stated, we always assume
that $f(t)\in C^{2}(R)$. 

In periodic case initial conditions (and the dynamics itself) are
in fact finite dimensional - one can assume that at time $0$ there
are exactly $N$ point particles $0,1,\ldots ,N-1$ with velocities $v_{0}(0),\ldots ,v_{N-1}(0)$
and coordinates inside $[0,L)$: 
\begin{equation}
0=x_{0}(0)<x_{1}(0)<\ldots <x_{N-1}(0)<x_{N}(0)=L\label{initial}
\end{equation}
To prove existence of periodic solution for all $t\in[0,\infty)$
is an easy matter, but very important problem arises: 

1) we call the dynamics (solution of equations) \textbf{regular} (without
collisions) if it cannot occur that $x_{k+1}(t)=x_{k}(t)$ for any
$k$ and $t\geq0$. Otherwise we call dynamics \textbf{irregular}.
In regular dynamics the order of particles is conserved. And it is
important to know for which parameters and initial conditions the
dynamics is regular. This problem is ignored in most papers (known
to us) on dynamics of linear chains (deterministic or random).

2) Another (piece-wise smooth) dynamics can be defined. Namely, when
the event (called \textbf{collision}) $x_{k}(t)=x_{k+1}(t)$ for some
$k$ and $t\geq0$ occurs, we assume that these particles exchange
velocities, and thus the order is conserved. Dynamics with collisions
is more complicated (than purely linear), but we consider such dynamics
as well. We do not consider multiple (for example, triple) collisions
because they could occur only for the set of initial conditions of
Lebesgue measure zero.

It is convenient to take $a=\frac{L}{N}$ in (\ref{potential}), because
then formally $U_{0}=0$ for $x_{k}=\frac{kL}{N}$, that does not
influence the main equations (\ref{x_equations}). For simpler presentation
it will be convenient to reduce this system to even simpler finite
dimensional. For this we introduce variables $q_{k}=x_{k+1}-x_{k}$.
Note that for any integer $m$ we have $q_{k}=q_{k+mN}$. They satisfy
the following $N$ equations 
\begin{equation}
\ddot{q}_{k}=\omega^{2}(q_{k+1}-2q_{k}+q_{k-1})-\alpha\dot{q}_{k}+f_{k+1}-f_{k}=\omega^{2}(q_{k+1}-2q_{k}+q_{k-1})-\alpha\dot{q}_{k},\label{q_equations}
\end{equation}
only on variables $q_{k},k=0,1,\ldots ,N-1$, as in these equations $q_{-1}=q_{N-1},q_{N}=q_{0}.$ 

Put
\[
Q(t)=\sum_{k=0}^{N-1}q_{k}(t),
\]
then 
\begin{equation}
Q(t)=Q(0)=L\label{Q}
\end{equation}
And we get that $\dot{Q}(t)\equiv0$ for all $t>0$. 

If we could find $q_{k}(t)$, then we can find $x_{k}(t)$ using,
for example, the equation
\begin{equation}
\ddot{x}_{k}=\omega^{2}(q_{k}(t)-q_{k-1}(t))-\alpha\dot{x}_{k}+f_{k},\label{x_q_equations}
\end{equation}
that is the one-particle equation with external driving force $F_{k}(t)=\omega^{2}(q_{k}(t)-q_{k-1}(t))+f_{k}$. 

Then we can identify the circle $S=S_{L}$ of length $L$ with the
segment $[0,L)$ with identified end points, and study the flow of
particles along this circle. Then there will be exactly $N$ particles
inside $[0,L)$ at any time $t>0$. When $q_{k}(t)\geq0$ for all
$k$ and $t$ ? For example, this will be even if we consider dynamics
with collisions. Important special cases, when $q_{k}(t)>0$ for all
$k$ and $t$, will be considered below.

Note that equations (\ref{q_equations}) can be considered as Hamiltonian
equations (with total energy $H=T+U$) plus dissipative forces , where
\begin{equation}
T=\sum_{k=0}^{N-1}\frac{\dot{q}_{k}^{2}}{2},\label{q_kinetic}
\end{equation}
\begin{equation}
U=U_{0}-\sum_{k=0}^{N-1}(f_{k+1}-f_{k})q_{k}=U_{0}=\frac{\omega^{2}}{2}\sum_{k=0}^{N-1}(q_{k+1}-q_{k})^{2}\label{q_potential}
\end{equation}

\section{Results}

Remind that further on we assume equal forces on all particles that
is $f_{k}\equiv f(t)$. 

\subsection{Convergence to uniform flow}

Here particles may collide and the collisions are elastic. Then
\begin{theorem}\label{const_forc} For any initial conditions (\ref{infinite}),
as $t\to\infty$
\[
q_{k}(t)=x_{k+1}(t)-x_{k}(t)\to\frac{L}{N}
\]
for all $k$. Moreover:

1) if $f(t)\equiv f=const$, then 
\[
\dot{x}_{k}(t)\to w=\frac{f}{\alpha},
\]

2) if $f(t)$ is periodic with period $2\pi$ and convergent Fourier
series
\begin{equation}
f(t)=\sum_{m\in Z}a_{m}e^{imt},\label{f_s}
\end{equation}
 then there exists periodic function with period $2\pi$
\[
w(t)=\sum_{m\in Z}\frac{a_{m}}{\alpha+im}e^{imt}
\]
such that for $t\to\infty$
\[
\dot{x}_{k}(t)-w(t)\to0
\]

3) if $f(t)$ is a second-order stationary process ($Ef^{2}(s)<\infty$)
with the continuous covariance function, finite mean value $\bar{f}=Ef(s)$
and orthogonal measure $\mu(du),$ i.e.
\begin{equation}
f(s)=\bar{f}+\int_{R}e^{isu}\mu(du),\label{spectral}
\end{equation}
then there exists the stationary process 
\[
w(t)=\frac{\bar{f}}{\alpha}+\int_{R}e^{itu}(\alpha+iu)^{-1}\mu(du)
\]
such that a.s. 
\[
\dot{x}_{k}(t)-w(t)\to0
\]
\end{theorem}

{\it Proof.} Consider equations (\ref{q_equations}) and corresponding kinetic
and potential energies defined by (\ref{q_kinetic}) and (\ref{q_potential}).
In case of equal forces we have $U=U_{0}.$ 

Between collisions the dynamics is given by equations (\ref{x_equations})
and (\ref{q_equations}). At the moment of collision the colliding
particles exchange velocities and, hence, the kinetic energy does
not change at any moment of collisions.

For dynamics, defined by equations (\ref{q_equations}), there exists
only one fixed point. It is easily found 
\[
\frac{\partial U_{0}}{\partial q_{k}}=0\Longleftrightarrow q_{k+1}=q_{k}=\frac{L}{N},\; k=0,1,\ldots ,N-1,
\]
because $\sum q_{k}\equiv L$. Moreover, the only minimum of potential
energy is reached at this point.

It is well-known and easy to check that 
\[
\frac{dH_{0}}{dt}=-\alpha\sum_{k=0}^{N-1}\dot{q}_{k}^{2}<0
\]
where $H_{0}=T+U_{0}.$ Indeed, using
\[
\ddot{q}_{k}=-\frac{\partial U_{0}}{\partial q_{k}}-\alpha\dot{q}_{k},\alpha>0,
\]
we get
\[
\frac{dH_{0}}{dt}=\sum_{k=0}^{N-1}\dot{q}_{k}\ddot{q}_{k}+\sum_{k=0}^{N-1}\frac{\partial U_{0}}{\partial q_{k}}\dot{q}_{k}=\sum_{k=0}^{N-1}\dot{q}_{k}(\ddot{q}_{k}+\frac{\partial U_{0}}{\partial q_{k}})=-\alpha\sum_{k=0}^{N-1}\dot{q}_{k}^{2}
\]
Note also that $H_{0}$ has the only minimum which is equal to $0$
and is reached at the point $\dot{q}_{k}=0,q_{k}=q_{k+1}=\frac{L}{N},k=0,\ldots ,N-1.$
This gives $\dot{q}_{k}(t)\to0$ and $T\to0$ as $t\to\infty$. Then
$U_{0}(t)$ should tend to its value at this point, that is to $0$. 

To get asymptotics of all $\dot{x}_{k}(t)$ for any initial conditions,
sum up equations (\ref{x_equations}). That gives for $X_{N}=\sum_{k=0}^{N-1}x_{k},V_{N}=\dot{X}_{N}$
the equation
\[
\ddot{X}_{N}=-\alpha\dot{X}_{N}+Nf(t)\Longleftrightarrow\dot{V}_{N}=-\alpha V_{N}+Nf(t)
\]
with the solution
\begin{equation}
V_{N}(t)=V_{N}(0)e^{-\alpha t}+N\int_{0}^{t}f(s)e^{-\alpha(t-s)}ds,\label{sol}
\end{equation}
If $f=const$ then $\dot{X}_{N}(t)$ converges to $\frac{Nf}{\alpha}$
and any $\dot{x}_{k}$ converges to $w=\frac{f}{\alpha}$. 

Let $f(t)$ be periodic. It follows from $\dot{q}_{k}(t)\to0$ that
all $\dot{x}_{k}(t)$ converge to the same function $w(t)$, for any
initial conditions. Then $\dot{X}_{N}(t)$ converges to $N\int_{0}^{t}f(s)e^{-\alpha(t-s)}ds$
and any $\dot{x}_{k}$ converges to $\int_{0}^{t}f(s)e^{-\alpha(t-s)}ds.$
By (\ref{f_s}) we have
\[
\int_{0}^{t}f(s)e^{-\alpha(t-s)}ds=\int_{0}^{t}\sum_{m\in Z}a_{m}e^{ims}e^{-\alpha(t-s)}ds=\sum_{m\in Z}a_{m}e^{-\alpha t}\int_{0}^{t}e^{(im+\alpha)s}ds=
\]
\[
=\sum_{m\in Z}\frac{a_{m}}{\alpha+im}e^{imt}-e^{-\alpha t}\sum_{m\in Z}\frac{a_{m}}{\alpha+im}
\]
So any $\dot{x}_{k}$ converges to 
\[
w(t)=\sum_{m\in Z}\frac{a_{m}}{\alpha+im}e^{imt}
\]

Let $f(t)$ be a stationary process. Then substituting (\ref{spectral})
into (\ref{sol}) we find
\[
V_{N}(t)=V_{N}(0)e^{-\alpha t}+N\int_{0}^{t}e^{-\alpha(t-s)}\bar{f}ds+N\int_{0}^{t}e^{-\alpha(t-s)}ds\int_{R}e^{isu}\mu(du)
\]
where the third term can be rewritten as
\[
N\int_{0}^{t}e^{-\alpha(t-s)}ds\int_{R}e^{isu}\mu(du)=Ne^{-\alpha t}\int_{R}\mu(du)\int_{0}^{t}e^{(\alpha+iu)s}ds=
\]
\[
=N\int_{R}e^{itu}(\alpha+iu)^{-1}\mu(du)-Ne^{-\alpha t}\int_{R}(\alpha+iu)^{-1}\mu(du)
\]
So we have
\[
V_{N}(t)=\frac{N\bar{f}}{\alpha}+N\int_{R}e^{itu}(\alpha+iu)^{-1}\mu(du)+O(e^{-\alpha t})
\]
and a.s.
\[
\dot{x}_{k}(t)-w(t)=\dot{x}_{k}(t)-\frac{\bar{f}}{\alpha}-\int_{R}e^{itu}(\alpha+iu)^{-1}\mu(du)\to0.
\]
The theorem is proved.

\subsection{Regularity conditions}

We would like to get at least sufficient conditions for there were
no collisions. Put
\[
\omega=\omega_{0}N
\]
and assume that $\omega_{0}>0,\alpha\geq0,L>0$ are fixed constants,
not depending on $N$. All further results will be for sufficiently
large $N$.

\paragraph{Initial conditions}

We shall say that periodic initial conditions have ``almost smooth
profiles'' if the following two conditions hold: 

1) there exist periodic functions $X(x),V(x)\in C^{4}(R)$ with period
$L$, where $X(x)>0$ for any $x\in R,$ and
\begin{equation}
\int_{0}^{L}X(u)du=L,\;\int_{0}^{L}V(u)du=0\label{z_m}
\end{equation}

2) for some constants $C_{1}>0,C_{2}>0$ 
\begin{equation}
  |x_{k+1}^{(N)}(0)-x_{k}^{(N)}(0)-\frac{L}{N}X(\frac{kL}{N})|<\frac{C_{1}}{N^{2}},\quad
  |\dot{x}_{k+1}^{(N)}(0)-\dot{x}_{k}^{(N)}(0)-\frac{L}{N}V(\frac{kL}{N})|<\frac{C_{2}}{N^{2}}\label{regularity_cond}
\end{equation}
uniformly in $k.$ 

Define constants
\begin{equation}
c_{1}=L\int_{0}^{L}|\frac{d^{2}X}{du^{2}}(u)|du,c_{2}=L\int_{0}^{L}|\frac{d^{2}V}{du^{2}}(u)|du,\label{c12}
\end{equation}
In some sense $c_{1},c_{2}$ define fluctuations of the ``profile''.
We will need also the constant 
\begin{equation}
\gamma=\gamma(X,V,\alpha,\omega_{0},C_{1},C_{2},c_{1},c_{2})=(1+\frac{\alpha}{8\omega_{0}})(2c_{1}+C_{1}L^{-1})+\frac{2c_{2}+C_{2}L^{-1}}{4\omega_{0}}>0\label{gamma}
\end{equation}
Let $\Omega_{N}^{(0)}(\delta)\subset R^{2N},0<\delta<1,$ be a set
of ``almost smooth'' initial conditions $x_{k}^{(N)}(0),\dot{x}_{k}^{(N)}(0),k=0,\ldots ,N-1,$
with additional condition that $\gamma(X,V)<\delta$. We shall prove
below the following

\begin{lemma}\label{Lemma_deviation-1}
For all $x\in R$
\begin{equation}
X(0)-c_{1}\leq X(x)\leq c_{1}+X(0)\label{deviation-1}
\end{equation}
\end{lemma}

Let $\Omega_{N}(\delta)$ be the domain of $R^{N}=\{(x_{0},\ldots ,x_{N-1})\}$,
defined for some $0<\delta<1$ by the estimates 
\[
|x_{k+1}-x_{k}-\frac{L}{N}|<\frac{L\delta}{N}
\]
 for all $k$. 

\begin{theorem}\label{absence_of_collisions}
Let initially the system belong to $\Omega_{N}^{(0)}(\delta)$ for
some $0<\delta<1.$ Then it stays in $\Omega_{N}(\delta)$ for all
$t\geq0$, that is 
\[
|x_{k+1}^{(N)}(t)-x_{k}^{(N)}(t)-\frac{L}{N}|<\frac{L\delta}{N}
\]
 for all $k,t$.
\end{theorem}

It follows that particles conserve the initial order at any time $t>0.$

\subsection{Convergence to regular continuum mechanics}

Concerning the term ``regular'' see \cite{LMCh}. This property
was ignored in many papers on mechanics of continuum media. But of
course not in all, see for example \cite{Marsden_1,Marsden_2,Marsden_3,Marsden_4}.

With each point $x\in R$ we associate the particle with number $k(x,N)$
such that 
\begin{equation}
x_{k(x,N)}^{(N)}(0)\leq x<x_{k(x,N)+1}^{(N)}(0)\label{k_x_N}
\end{equation}

\begin{theorem}\label{wave}
Under conditions of theorem \ref{absence_of_collisions} we have

1) For any $T>0$ uniformly in $t\in[0,T]$ and in $x\in R$ there
exists the limit 
\begin{equation}
\lim_{N\to\infty}x_{k(x,N)}^{(N)}(t)=Y(t,x)\in R\label{w2}
\end{equation}
where function $Y(t,x)$ satisfies the condition $Y(t,x+L)=Y(t,x)+L$
for any $x\in R.$

2) Moreover, $Y(t,x):\,R\to R$ is differentiable in $x$ and $t$
and strictly increasing in $x$ for each fixed $t$. So it is a diffeomorphism
of $R$ for any $t$ . 
\end{theorem}

The function $Y(t,x)\in R$ will be called the trajectory of the continuous
media particle which is initially at point $x\in R.$ 

Define for $x\in R$
\[
\pi(x)=x,\mod L
\]
So $\pi:\,R\to S^{1}=[0,L).$ One can define the trajectory $y(t,x)\in S^{1}$
of the point $x$ on the circle by the equation
\[
\pi(Y(t,x))=y(t,\pi(x)),\; x\in R
\]
Also the mapping $y(t,s):\,S^{1}\to S^{1}$ is a diffeomorphism of
the circle.

For given $N$ define the distribution function on $[0,L)$
\[
F^{(N)}(t,y)=\frac{1}{N}\sharp\{k\in\{0,1,\ldots ,N-1\}:\pi(x_{k}^{(N)}(t))\leq y\},\; y\in[0,L)
\]

Let $x(t,y)\in[0,L)$ be the map inverse to $y(t,x)$, that is $y(t,x(t,y))=y$.
This map exists according to theorem \ref{wave}. Introduce the function
$z(x):R\to R$ by the equation:
\begin{equation}
\int_{0}^{z(x)}X(x^{\prime})dx^{\prime}=x\label{Z_x}
\end{equation}
The inverse function $x(z)$, that is such that $x(z(x))=x$. 

\begin{lemma}\label{density}Uniformly in $y\in[0,L)$ and in $t\in[0,T]$,
for any $T<\infty$, we have
\begin{equation}
\lim_{N\to\infty}F^{(N)}(t,y)=F(t,y)=\frac{z(x(t,y))}{L},y\in[0,L),\label{distr_fun}
\end{equation}
where $F(t,y)$ is twice differentiable in $y$ and $t$. 
\end{lemma}

Define the density of ``the number of continuum media particles''
as
\begin{equation}
\rho(t,y)=\frac{dF(t,y)}{dy},\;\, y\in[0,L)\label{den}
\end{equation}

As the particles do not collide, then one can unambiguously define
the function $u(t,y)$ as the speed of the (unique) particle situated
at time $t$ at the point $y$, that is 
\[
u(t,y(t,x))=\frac{dy(t,x)}{dt}.
\]
For $t=0$
\[
F(0,x)=\frac{z(x)}{L}\leq1,\quad  \rho(0,x)=\frac{z^{\prime}(x)}{L}=\frac{1}{LX(z(x))}
\]

\subsection{Oscillator chain and wave equation}

In many textbooks it is said that, under some scaling, dynamics of
oscillator chain with $N$ oscillators converges to one-dimensional
wave equation if $N\to\infty$. But it appears, as we shall see now,
that one should be more care -- this strongly depends on the choice
of space variable and on the initial conditions.

We define the function $G(t,z):R\to R$ 
\begin{equation}
G(t,z)=Y(t,x(z))\Longleftrightarrow Y(t,x)=G(t,z(x))\label{Y_G}
\end{equation}

\begin{theorem} \label{wave_eq}Let $\omega_{1}=\omega_{0}L$.

1) The equation for $Y(t,x(z))$ is
\[
Y_{tt}(t,x(z))=\omega_{1}^{2}\left(Y_{xx}(t,x(z))X^{2}(z)+Y_{x}(t,x(z))X^{\prime}(z)\right)-\alpha Y_{t}(t,x(z))+f(t)
\]
with initial conditions
\[
Y(0,x(z))=x(z)=\int_{0}^{z}X(u)du=G(0,z)
\]
\[
Y_{t}(0,x(z))=v+\int_{0}^{z}V(u)du
\]

2) The function $G(t,z)$ satisfies the inhomogeneous wave equation
\begin{equation}
G_{tt}(t,z)=\omega_{1}^{2}G_{zz}(t,z)-\alpha G_{t}(t,z)+f(t)\label{Wave}
\end{equation}
\end{theorem}

It follows that if $\alpha=0,f=0$ then the limiting equation is
\begin{equation}
Y_{tt}(t,x(z))=\omega_{1}^{2}\left(Y_{xx}(t,x(z))X^{2}(z)+Y_{x}(t,x(z))X^{\prime}(z)\right),\label{classical}
\end{equation}
where $\omega_{1}=\omega_{0}L$. It becomes classical
\[
Y_{tt}(t,x)=\omega_{1}^{2}Y_{xx}(t,x)
\]
 only if $X(x)=1$. 

\subsection{Explicit dynamics in Lagrange coordinates}

We start with the case $\alpha=f=0$. Consider the homogeneous wave
equation
\[
G_{tt}(t,z)=\omega_{1}^{2}G_{zz}(t,z),\:z\in R
\]
with initial conditions
\begin{equation}
G(0,z)=\phi(z)=\int_{0}^{z}X(u)du,\;G_{t}(0,z)=\psi(z)=v+\int_{0}^{z}V(u)du,v=\dot{x}_{0}(0)\label{I_C}
\end{equation}
Note that
\[
\phi(z+L)=\phi(z)+L,\psi(z+L)=\psi(z)
\]
The d'Alembert solution can be written as follows
\begin{equation}
G(t,z)=\frac{1}{2}(\phi(z+\omega_{1}t)+\phi(z-\omega_{1}t))+\frac{1}{2\omega_{1}}\int_{z-\omega_{1}t}^{z+\omega_{1}t}\psi(y)dy\label{waves-1}
\end{equation}
One can also write this solution in the form
\[
G(t,z)=G_{+}(z+\omega_{1}t)+G_{-}(z-\omega_{1}t)
\]
where 
\[
G_{\pm}(z)=\frac{1}{2}(\phi(z)\pm\frac{1}{\omega_{1}}\int_{0}^{z}\psi(y)dy+C_{\pm})
\]
and constants $C_{\pm}$ satisfy condition $C_{+}+C_{-}=0.$ Note
that $G(t,z+L)=G(t,z)+L.$ Then $Y(t,x)$ is given by (\ref{Y_G}).

Using d'Alembert solution (\ref{waves-1}) one can easily get the
following lemma.

\begin{lemma} Let $\alpha=f=0.$ Then for any fixed $t$ the function
$G(t,z):R\to R$ is a diffeomorphism if for all $z\in R$
\[
G_{z}(t,z)=X(z+\omega_{1}t)+X(z-\omega_{1}t)+\frac{1}{\omega_{1}}\int_{z-\omega_{1}t}^{z+\omega_{1}t}V(y)dy>0
\]
\end{lemma}

For arbitrary $\alpha>0$ and $f$ we prove below the following result.

\begin{theorem}\label{exp_dyn}
  $\phantom{a}$
  
1) The solution of (\ref{Wave}) with initial conditions (\ref{I_C})
is
\[
G(t,z)=\frac{e^{-\frac{\alpha}{2}t}}{2}(\phi(z+\omega_{1}t)+\phi(z-\omega_{1}t))+
\]
\[
+\frac{\alpha e^{-\frac{\alpha}{2}t}}{4\omega_{1}}\int_{z-\omega_{1}t}^{z+\omega_{1}t}\Bigl(t\frac{I_{1}(\frac{\alpha}{2}\sqrt{t^{2}-(z-\xi)^{2}/\omega_{1}^{2}})}{\sqrt{t^{2}-(z-\xi)^{2}/\omega_{1}^{2}}}+I_{0}(\frac{\alpha}{2}\sqrt{t^{2}-(z-\xi)^{2}/\omega_{1}^{2}})\Bigr)\phi(\xi)d\xi+
\]
\[
+\frac{e^{-\frac{\alpha}{2}t}}{2\omega_{1}}\int_{z-\omega_{1}t}^{z+\omega_{1}t}I_{0}(\frac{\alpha}{2}\sqrt{t^{2}-(z-\xi)^{2}/\omega_{1}^{2}})\psi(\xi)d\xi+
\]
\[
+\frac{1}{2\omega_{1}}\int_{0}^{t}e^{-\frac{\alpha}{2}(t-\tau)}f(\tau)d\tau\int_{z-\omega_{1}(t-\tau)}^{z+\omega_{1}(t-\tau)}I_{0}(\frac{\alpha}{2}\sqrt{(t-\tau)^{2}-(z-\xi)^{2}/\omega_{1}^{2}})d\xi,
\]
where $I_{0}(x),I_{1}(x)$ are modified Bessel functions:
\[
I_{0}(x)=\sum_{m=0}^{\infty}\frac{1}{m!\Gamma(m+1)}\left(\frac{x}{2}\right)^{2m}
\]
\[
I_{1}(x)=\sum_{m=0}^{\infty}\frac{1}{m!\Gamma(m+2)}\left(\frac{x}{2}\right)^{2m+1}
\]

2) For any fixed $t$ the function $G(t,z):R\to R$ is a diffeomorphism
if $G_{z}(t,z)>0$ for all $z\in R.$
\end{theorem}

\subsection{Conservation law, Euler equation and pressure}

\begin{theorem}\label{euler}
Let $\omega_{1}=\omega_{0}L$ and the conditions of the theorem \ref{absence_of_collisions}
hold. For any $t>0$, $y\in[0,L)$ we have :
\begin{equation}
\frac{\partial\rho(t,y)}{\partial t}+\frac{d}{dy}(u(t,y)\rho(t,y))=0\label{cons_law}
\end{equation}
\begin{align}
  \frac{\partial u(t,y)}{\partial t}+u(t,y)\frac{\partial u(t,y)}{\partial y}+\alpha u(t,y)-f(t)&=
                                                                                                  -\frac{\omega_{1}^{2}\rho_{y}(t,y)}{\rho^{3}(t,y)}
                                                                                                  =\frac{1}{\rho(t,y)}\frac{d}{dy}\frac{\omega_{1}^{2}}{\rho(t,y)} \notag \\
  &=-\frac{p_{y}(t,y)}{\rho(t,y)}\label{Euler}
\end{align}
 where $p(t,y)$ is called pressure and is defined as follows:
\begin{equation}
p(t,y)=-\frac{\omega_{1}^{2}}{\rho(t,y)}+C\label{pressure}
\end{equation}
for some constant $C$. 
\end{theorem}

Constant $C$ can be chosen as $C=\omega_{1}^{2}$ , so that at equilibrium
(when $\rho=1$) the pressure were zero.

For given $y$ and $t$ define the number $k(y,N,t)$ so that 
\[
x_{k(y,N,t)}^{(N)}(t)\leq y<x_{k(y,N,t)+1}^{(N)}(t)
\]
Consider the point $y\in[0,L)$ and the force acting on the particle
with number $k(y,N,t)$:
\begin{align}
  R^{(N)}(t,y)&=\omega^{2}(x_{k(y,N,t)+1}^{(N)}(t)-x_{k(y,N,t)}^{(N)}(t)-\frac{L}{N})\notag \\
  &\quad {} -\omega^{2}(x_{k(y,N,t)}^{(N)}(t)-x_{k(y,N,t)-1}^{(N)}(t)-\frac{L}{N})\label{r_n}
\end{align}

\begin{theorem}\label{force}
Let the conditions of the theorem \ref{absence_of_collisions} hold.
Then for any $0<T<\infty$, uniformly in $y\in[0,L)$ and in $t\in[0,T]$
the following limit exists:
\begin{equation}
\lim_{N\rightarrow\infty}R^{(N)}(t,y)=-\frac{p_{y}(t,y)}{\rho(t,y)},\label{interaction_strength}
\end{equation}
where the functions $p,\rho$ are the same as in theorem \ref{euler}.
\end{theorem}

Note that as the pressure is defined up to an additive constant, it
can be considered as an ``interaction potential'' for continuum
media, an analog of interaction potentials in Hamiltonian particle
mechanics.

\subsection{Euler equations in Lagrangian coordinates}

Consider the density and the velocity in Lagrangian coordinates 
\begin{equation}
\hat{\rho}(t,z)=L\rho(t,y(t,x(z)))=L\rho(t,G(t,z))\label{first}
\end{equation}
\begin{equation}
\hat{u}(t,z)=\frac{\partial y(t,x(z))}{\partial t}=\frac{\partial G(t,z)}{\partial t}\label{second}
\end{equation}
\begin{theorem}\label{Euler_Lagrange} Let the conditions of the
theorem \ref{absence_of_collisions} be satisfied. For any $t>0,z\in[0,L)$
we have
\begin{equation}
\frac{\partial}{\partial t}\left(\frac{1}{\hat{\rho}(t,z)}\right)-\frac{\partial\hat{u}(t,z)}{\partial z}=0\label{fe}
\end{equation}
\begin{equation}
\frac{\partial\hat{u}(t,z)}{\partial t}+\alpha\hat{u}(t,z)-f(t)=-\frac{\partial\hat{p}(t,z)}{\partial z}=\frac{\partial}{\partial t}\left(\frac{\omega_{1}^{2}}{\hat{\rho}(t,z)}\right)\label{se}
\end{equation}
where 
\[
\hat{p}(t,z)=-\frac{\omega_{1}^{2}}{\hat{\rho}(t,z)}+C
\]
\end{theorem}

\section{Proofs}

\subsection{Proof of Lemma \ref{Lemma_deviation-1}}

We shall use the following simple assertion. Assume that $f(x)\in C^{2}(R)$
is periodic with period $L$ and $f(0)=0$. Then the following inequality
holds
\[
\sup_{x\in R}|f(x)|\leq L\int_{0}^{L}|f^{\prime\prime}(x)|dx.
\]

Indeed, 
\[
f(x)=\int_{0}^{x}f^{\prime}(u)du,x\in[0,L]
\]
 It follows that 
\[
\sup_{x\in R}|f(x)|\leq\int_{0}^{L}|f^{\prime}(u)|du.
\]
 Dince $f(0)=f(L)=0$,  there exists a point $x^{*}\in(0,L)$ such
that $f^{\prime}(x^{*})=0$. Thus, we have
\begin{align*}
  \int_{0}^{L}|f^{\prime}(x)|dx&=\int_{0}^{L}\Biggl|\int_{x}^{x^{*}}f^{\prime\prime}(u)du\Biggr| dx\\
  &\leq
  L\sup_{x\in[0,L]}\Biggl| \int_{x}^{x^{*}}f^{\prime\prime}(u)du\Biggr| \leq L\int_{0}^{L}|f^{\prime\prime}(u)|du
\end{align*}
 and the assertion follows. To prove the Lemma put $f(x)=X(x)-X(0).$
Then for all $x\in R$ 
\[
X(0)-c_{1}\leq X(x)\leq X(0)+c_{1}
\]
where 
\[
c_{1}=L\int_{0}^{L}|X^{\prime\prime}(u)|du
\]

\subsection{Proof of theorem \ref{absence_of_collisions}}

Define variables $r_{k}^{(N)}(t)$ as follows
\[
r_{k}^{(N)}(t)=q_{k}^{(N)}(t)-\frac{L}{N}=x_{k+1}^{(N)}(t)-x_{k}^{(N)}(t)-\frac{L}{N}
\]
Further we shall omit the upper index $N$ for simplicity. Then
\[
\dot{r}_{k}(t)=\dot{q}_{k}(t)=\dot{x}_{k+1}(t)-\dot{x}_{k}(t)
\]
It follows from (\ref{q_equations}) that variables $r_{k}$ satisfy
the system 
\begin{equation}
\ddot{r}_{k}=\omega^{2}(r_{k+1}-2r_{k}+r_{k-1})-\alpha\dot{r}_{k}\label{r_equations}
\end{equation}
where $k=0,1,\ldots ,N-1,$ and $r_{-1}=r_{N-1},r_{N}=r_{0}.$ 

By (\ref{Q}) we have
\[
R_{0}(t)=\sum_{k=0}^{N-1}r_{k}(t)\equiv0,\;\dot{R}_{0}(t)\equiv0
\]
By (\ref{regularity_cond}) 
\[
|r_{k}(0)-\frac{L}{N}(X(\frac{kL}{N})-1)|<\frac{C_{1}}{N^{2}},\quad |\dot{r}_{k}(0)-\frac{L}{N}V(\frac{kL}{N})|<\frac{C_{2}}{N^{2}}
\]

We will use the discrete Fourier transform
\begin{equation}
R_{j}(t)=\sum_{k=0}^{N-1}r_{k}(t)e^{i\frac{2\pi jk}{N}},\quad r_{k}(t)=\frac{1}{N}\sum_{j=0}^{N-1}R_{j}(t)e^{-i\frac{2\pi jk}{N}}\label{fourier_transform}
\end{equation}
\begin{equation}
\dot{R}_{j}(t)=\sum_{k=0}^{N-1}\dot{r}_{k}(t)e^{i\frac{2\pi jk}{N}},\quad \dot{r}_{k}(t)=\frac{1}{N}\sum_{j=0}^{N-1}\dot{R}_{j}(t)e^{-i\frac{2\pi jk}{N}}\label{der_four_trans}
\end{equation}
Calculating the discrete Fourier transform of both parts
of (\ref{r_equations}), we obtain a system of decoupled differential
equations for Fourier images $R_{j}:$
\begin{equation}
\ddot{R}_{j}=-\Omega_{j}^{2}R_{j}-\alpha\dot{R}_{j}\label{rj}
\end{equation}
for $j=1,\ldots ,N-1,$ where
\[
\Omega_{j}^{2}=2\omega^{2}(1-\cos(\frac{2\pi j}{N}))=4\omega^{2}\sin^{2}(\frac{\pi j}{N}),\quad j=1,\ldots ,N-1\Longleftrightarrow
\]
\[
\Omega_{j}=2\omega\sin(\frac{\pi j}{N})=2\omega_{0}N\sin(\frac{\pi j}{N})>0.
\]
Note that
\[
\Omega_{j}=\Omega_{N-j}.
\]
Using the inequality
\[
\frac{2}{\pi}<\frac{\sin x}{x}<1,\quad 0<x<\frac{\pi}{2}
\]
we get
\[
\frac{2j}{N}<\sin(\frac{\pi j}{N})<\frac{\pi j}{N}\Longleftrightarrow
\]
\begin{equation}
4\omega_{0}j<\Omega_{j}<2\pi\omega_{0}j,\:j=1,\ldots ,[N/2],\label{sigma_j}
\end{equation}
\[
4\omega_{0}j<\Omega_{N-j}<2\pi\omega_{0}j,\:j=1,\ldots ,[N/2]
\]

Find roots of the quadratic equation
\[
\lambda^{2}+\alpha\lambda+\Omega_{j}^{2}=0,\quad j=1,\ldots ,N-1
\]
They equal
\[
\lambda_{1,2}(j)=-\frac{\alpha}{2}\pm d_{j},\;\frac{\alpha^{2}}{4}-\Omega_{j}^{2}\geq0
\]
\[
\lambda_{1,2}(j)=-\frac{\alpha}{2}\pm id_{j},\;\frac{\alpha^{2}}{4}-\Omega_{j}^{2}<0
\]
where 
\begin{equation}
d_{j}=\sqrt{\left|\frac{\alpha^{2}}{4}-\Omega_{j}^{2}\right|}=\sqrt{\left|\frac{\alpha^{2}}{4}-4\omega^{2}\sin^{2}(\frac{\pi j}{N})\right|},\quad j=1,\ldots ,N-1\label{d_j}
\end{equation}
The solution of equation (\ref{rj}) with initial conditions $R_{j}(0),\dot{R}_{j}(0)$
has the form 
\begin{equation}
R_{j}(t)=e^{-\alpha t/2}((\cos d_{j}t+\frac{\alpha\sin(d_{j}t)}{2d_{j}})R_{j}(0)+\frac{\sin(d_{j}t)}{d_{j}}\dot{R}_{j}(0))\label{R_j_complex}
\end{equation}
in case of $\frac{\alpha^{2}}{4}-\Omega_{j}^{2}<0$. In particular,
for $\alpha=0$ we have
\[
R_{j}(t)=\cos(\Omega_{j}t)R_{j}(0)+\frac{\sin(\Omega_{j}t)}{\Omega_{j}}\dot{R}_{j}(0)
\]

If $\frac{\alpha^{2}}{4}-\Omega_{j}^{2}>0$, then the solution is
equal to
\begin{equation}
R_{j}(t)=e^{-\alpha t/2}((\cosh d_{j}t+\frac{\alpha\sinh(d_{j}t)}{2d_{j}})R_{j}(0)+\frac{\sinh(d_{j}t)}{d_{j}}\dot{R}_{j}(0))\label{R_j_real}
\end{equation}
If $\frac{\alpha^{2}}{4}-\Omega_{j}^{2}=0,$ then
\begin{equation}
R_{j}(t)=e^{-\alpha t/2}((1+\frac{\alpha t}{2})R_{j}(0)+t\dot{R}_{j}(0))\label{R_j_multiple}
\end{equation}
To simplify notation introduce the following functions
\begin{equation}
a(d_{j}t)=\begin{cases}
e^{-\alpha t/2}\cosh(d_{j}t) & \frac{\alpha^{2}}{4}-\Omega_{j}^{2}\geq0\\
e^{-\alpha t/2}\cos(d_{j}t) & \frac{\alpha^{2}}{4}-\Omega_{j}^{2}<0
\end{cases}\label{fun_a}
\end{equation}
\begin{equation}
b(d_{j}t)=\begin{cases}
e^{-\alpha t/2}\frac{\sinh(d_{j}t)}{d_{j}} & \frac{\alpha^{2}}{4}-\Omega_{j}^{2}>0\\
e^{-\alpha t/2}\frac{\sin(d_{j}t)}{d_{j}} & \frac{\alpha^{2}}{4}-\Omega_{j}^{2}<0\\
e^{-\alpha t/2}t & \frac{\alpha^{2}}{4}-\Omega_{j}^{2}=0
\end{cases}\label{fun_b}
\end{equation}
In this notation the solution of (\ref{rj}) has the form
\begin{equation}
R_{j}(t)=(a(d_{j}t)+\frac{\alpha b(d_{j}t)}{2})R_{j}(0)+b(d_{j}t)\dot{R}_{j}(0),\quad j=1,\ldots ,N-1\label{R_j_general}
\end{equation}
 Applying the inverse Fourier transform
\begin{equation}
r_{k}(t)=\frac{1}{N}\sum_{j=1}^{N-1}R_{j}(t)e^{-i\frac{2\pi jk}{N}}\label{dev}
\end{equation}
 we find
\begin{equation}
r_{k}(t)=\frac{1}{N}\sum_{j=1}^{N-1}((a(d_{j}t)+\frac{\alpha b(d_{j}t)}{2})R_{j}(0)+b(d_{j}t)\dot{R}_{j}(0))e^{-i\frac{2\pi jk}{N}}\label{inverse_fourier_transform}
\end{equation}

By (\ref{inverse_fourier_transform}) 
\[
|r_{k}(t)|\leq\frac{1}{N}\sum_{j=1}^{N-1}((|a(d_{j}t)|+\frac{\alpha|b(d_{j}t)|}{2})|R_{j}(0)|+|b(d_{j}t)||\dot{R}_{j}(0)|)
\]
Note that
\[
|a(d_{j}t)|\leq1,\quad |b(d_{j}t)|\leq\frac{1}{\Omega_{j}}
\]
for $j=1,\ldots ,N-1.$ 

Hence,
\begin{equation}
|r_{k}(t)|\leq\frac{1}{N}\sum_{j=1}^{N-1}(1+\frac{\alpha}{2\Omega_{j}})|R_{j}(0)|+\frac{1}{N}\sum_{j=1}^{N-1}\frac{|\dot{R}_{j}(0)|}{\Omega_{j}}\label{rk}
\end{equation}
By inequality (\ref{sigma_j}) we have $\Omega_{j}\geq4\omega_{0}.$
So
\[
|r_{k}(t)|\leq(1+\frac{\alpha}{8\omega_{0}})\frac{1}{N}\sum_{j=1}^{N-1}|R_{j}(0)|+\frac{1}{4\omega_{0}N}\sum_{j=1}^{N-1}|\dot{R}_{j}(0)|
\]

Remind that 
\[
r_{k}(0)=x_{k+1}(0)-x_{k}(0)-\frac{L}{N}=\frac{L}{N}(X(\frac{kL}{N})-1)+\xi_{k,N}
\]
where we put $\xi_{k,N}=x_{k+1}(0)-x_{k}(0)-\frac{L}{N}X(\frac{kL}{N}).$
So for $j\neq0$
\[
R_{j}(0)=\frac{L}{N}\sum_{k=0}^{N-1}(X(\frac{kL}{N})-1)e^{i\frac{2\pi jk}{N}}+\sum_{k=0}^{N-1}\xi_{k,N}e^{i\frac{2\pi jk}{N}}=
\]
\[
=\frac{L}{N}\sum_{k=0}^{N-1}X(\frac{kL}{N})e^{i\frac{2\pi jk}{N}}+\sum_{k=0}^{N-1}\xi_{k,N}e^{i\frac{2\pi jk}{N}}
\]
because for $j\neq0$
\[
\sum_{k=0}^{N-1}e^{i\frac{2\pi jk}{N}}=0
\]
By condition (\ref{regularity_cond})
\begin{equation}
\left|\sum_{k=0}^{N-1}\xi_{k,N}e^{i\frac{2\pi jk}{N}}\right|\leq\sum_{k=0}^{N-1}|\xi_{k,N}|\leq\frac{C_{1}}{N}\label{one}
\end{equation}
As 
\[
\frac{L}{N}\sum_{k=0}^{N-1}X(\frac{kL}{N})e^{i\frac{2\pi jk}{N}}
\]
is the integral sum corresponding to the integral
\[
\int_{0}^{L}X(s)e^{i\frac{2\pi js}{L}}ds,
\]
we have
\begin{equation}
\left|\frac{L}{N}\sum_{k=0}^{N-1}X(\frac{kL}{N})e^{i\frac{2\pi jkL}{LN}}-\int_{0}^{L}X(s)e^{i\frac{2\pi js}{L}}ds\right|\leq\frac{c^{\prime}}{N}\label{two}
\end{equation}
As $X\in C^{2}$, we have for Fourier coefficients 
\[
\hat{X}_{j}=\int_{0}^{L}X(s)e^{i\frac{2\pi js}{L}}ds,
\]
corresponding to the function $X(s)$, the well known estimate
\[
|\hat{X}_{j}|\leq\frac{L^{2}\int_{0}^{L}|\ddot{X}(s)|ds}{j^{2}}=\frac{Lc_{1}}{j^{2}}
\]
So
\begin{equation}
\left|\int_{0}^{L}X(s)e^{i\frac{2\pi js}{L}}ds\right|\leq\frac{Lc_{1}}{j^{2}}\label{three}
\end{equation}
Thus, by (\ref{one}), (\ref{two}), (\ref{three}) we get
\begin{align}
  &\frac{1}{N}\sum_{j=1}^{N-1}|R_{j}(0)|\notag \\
  &\leq\frac{1}{N}\sum_{j=1}^{N-1}(|\hat{X}_{j}|+|\sum_{k=0}^{N-1}\xi_{k,N}e^{i\frac{2\pi jk}{N}}|+|\hat{X}_{j}-\frac{L}{N}\sum_{k=0}^{N-1}X(\frac{kL}{N})e^{i\frac{2\pi jkL}{LN}}|)\notag \\
  &\leq\frac{2Lc_{1}+C_{1}}{N}+O(N^{-2})\label{rk_1}
\end{align}
as $\sum_{j=1}^{\infty}j^{-2}<2.$

Similar, one can prove
\begin{equation}
\frac{1}{N}\sum_{j=1}^{N-1}|\dot{R}_{j}(0)|\leq\frac{2Lc_{2}+C_{2}}{N}+O(N^{-2})\label{rk_2}
\end{equation}
Indeed,
\[
\dot{r}_{k}(0)=\dot{x}_{k+1}(0)-\dot{x}_{k}(0)=\frac{L}{N}V(\frac{kL}{N})+\eta_{k,N}
\]
where we put $\eta_{k,N}=\dot{x}_{k+1}(0)-\dot{x}_{k}(0)-\frac{L}{N}V(\frac{kL}{N}).$
So for $j\neq0$
\[
\dot{R}_{j}(0)=\frac{L}{N}\sum_{k=0}^{N-1}V(\frac{kL}{N})e^{i\frac{2\pi jk}{N}}+\sum_{k=0}^{N-1}\eta_{k,N}e^{i\frac{2\pi jk}{N}}
\]
where 
\[
\left|\sum_{k=0}^{N-1}\eta_{k,N}e^{i\frac{2\pi jk}{N}}\right|\leq\sum_{k=0}^{N-1}|\eta_{k,N}|\leq\frac{C_{2}}{N}
\]
Similarly to (\ref{rk_1}) one can write the estimate
\begin{align*}
  \frac{1}{N}\sum_{j=1}^{N-1}|\dot{R}_{j}(0)|&\leq\frac{1}{N}\sum_{j=1}^{N-1}(|\hat{V}_{j}|+|\sum_{k=0}^{N-1}\eta_{k,N}e^{i\frac{2\pi jk}{N}}|+|\hat{V}_{j}-\frac{L}{N}\sum_{k=0}^{N-1}V(\frac{kL}{N})e^{i\frac{2\pi jkL}{LN}}|)\\
  &\leq\frac{2Lc_{2}+C_{2}}{N}+O(N^{-2})
\end{align*}
where
\[
\hat{V}_{j}=\int_{0}^{L}V(s)e^{i\frac{2\pi js}{L}}ds
\]

To finish the proof of the theorem note that by (\ref{rk_1}), (\ref{rk_2})
we have for $k=0,1,\ldots ,N-1$ uniformly over $t$ 
\[
|r_{k}(t)|\leq\frac{L\gamma}{N}+O(N^{-2})\leq\frac{L\delta}{N}
\]
for $\gamma<\delta$ and for sufficiently large $N.$ It follows that
\[
\frac{L(1-\delta)}{N}\leq x_{k+1}^{(N)}(t)-x_{k}^{(N)}(t)\leq\frac{L(1+\delta)}{N}
\]
For $\delta<1$ we get $x_{k+1}^{(N)}(t)-x_{k}^{(N)}(t)>0$ for sufficiently
large $N.$ So the initial order of particles is conserved for all
$t>0.$
The theorem is proved.

\subsection{Proof of theorem \ref{wave}}

Plan of the proof: 

1) We prove that
\[
\lim_{N\to\infty}x_{[\frac{zN}{L}]}^{(N)}(t)=G(t,z),\:z\in R,
\]
where $g(t,z)$ is the solution of the nonhomogeneous wave equation
with dissipation (\ref{nonhom_eq}) (see below).

2) Then we show
\[
\lim_{N\to\infty}x_{k(x,N)}^{(N)}(t)=G(t,z(x)),\:x\in R,
\]
 where the function $z(x):\:R\to R$ is uniquely defined by the equation:
\[
\int_{0}^{z(x)}X(x^{\prime})dx^{\prime}=x
\]

3) Finally, we define the trajectory $Y(t,x)=G(t,z(x)),x\in R.$ 

\subsubsection{Wave equations with dissipation}

Let $\omega_{1}=\omega_{0}L.$ Consider the homogeneous wave equation
with dissipation:
\begin{equation}
r_{tt}(t,x)=-\alpha r_{t}(t,x)+\omega_{1}^{2}r_{xx}(t,x),\; x\in R\label{hom_eq}
\end{equation}
We assume also the initial conditions
\begin{equation}
r(0,x)=\hat{Y}(x)=X(x)-1,r_{t}(0,x)=V(x),\; x\in R\label{in_con}
\end{equation}
and the periodic boundary condition
\[
r(t,x)=r(t,x+L),\; x\in R
\]
Let $\psi(x)$ be a periodic function with period $L$ and with zero
mean value. Define Fourier coefficients
\[
\hat{\psi}_{n}=\frac{1}{L}\int_{0}^{L}e^{\frac{i2\pi nx}{L}}\psi(x)dx
\]
Note that $\hat{\psi}_{0}=0.$ So we have
\[
\psi(x)=\sum_{n:n\neq0}\hat{\psi}_{n}e^{-\frac{i2\pi nx}{L}}
\]
Put
\begin{equation}
\lambda_{n}=\frac{2\pi n}{L},\quad \nu_{n}=\sqrt{|\lambda_{n}^{2}\omega_{1}^{2}-\alpha^{2}/4|}=\sqrt{|(2\pi n\omega_{0})^{2}-\alpha^{2}/4|}\label{nu}
\end{equation}
and introduce operator $\hat{G}(t)$ acting on periodic functions
$\psi$ as
\begin{equation}
\hat{G}(t)\psi=\sum_{n:n\neq0}\hat{\psi}_{n}e^{-i\lambda_{n}x}b(t\nu_{n})\label{G}
\end{equation}

\begin{lemma}\label{le1}There exists the only periodic solution
of (\ref{hom_eq}) with initial conditions (\ref{in_con}). This solution
has the form
\begin{equation}
r(t,x)=\frac{\partial(\hat{G}(t)\hat{Y})}{\partial t}+\alpha\hat{G}(t)\hat{Y}+\hat{G}(t)V\label{r}
\end{equation}
\end{lemma}

One can rewrite this solution in the explicit form
\begin{equation}
r(t,x)=\sum_{n:n\neq0}\hat{X}_{n}e^{-i\lambda_{n}x}(\tilde{a}(t\nu_{n})+\frac{\alpha\tilde{b}(t\nu_{n})}{2})+\sum_{n:n\neq0}\hat{V}_{n}e^{-i\lambda_{n}x}\tilde{b}(t\nu_{n})\label{sol_hom_eq}
\end{equation}
where 
\[
\hat{X}_{n}=\frac{1}{L}\int_{0}^{L}e^{\frac{i2\pi nx}{L}}(X(x)-1)dx=\frac{1}{L}\int_{0}^{L}e^{\frac{i2\pi nx}{L}}X(x)dx,\quad n\neq0
\]
\[
\hat{V}_{n}=\frac{1}{L}\int_{0}^{L}e^{\frac{i2\pi nx}{L}}V(x)dx
\]
and functions $\tilde{a},\tilde{b}$ are defined by 
\begin{equation}
\tilde{a}(\nu_{n}t)=\begin{cases}
e^{-\alpha t/2}\cosh(\nu_{n}t) & \frac{\alpha^{2}}{4}-\lambda_{n}^{2}\omega_{1}^{2}\geq0\\
e^{-\alpha t/2}\cos(\nu_{n}t) & \frac{\alpha^{2}}{4}-\lambda_{n}^{2}\omega_{1}^{2}<0
\end{cases}\label{fun_a-1}
\end{equation}
\begin{equation}
\tilde{b}(\nu_{n}t)=\begin{cases}
e^{-\alpha t/2}\frac{\sinh(\nu_{n}t)}{\nu_{n}} & \frac{\alpha^{2}}{4}-\lambda_{n}^{2}\omega_{1}^{2}>0\\
e^{-\alpha t/2}\frac{\sin(\nu_{n}t)}{\nu_{n}} & \frac{\alpha^{2}}{4}-\lambda_{n}^{2}\omega_{1}^{2}<0\\
e^{-\alpha t/2}t & \frac{\alpha^{2}}{4}-\lambda_{n}^{2}\omega_{1}^{2}=0
\end{cases}\label{fun_b-1}
\end{equation}

\paragraph{Proof of lemma \ref{le1}}

We can represent the solution by Fourier series
\[
r(t,x)=\sum_{n\in Z}\hat{r}_{n}(t)e^{-\frac{i2\pi nx}{L}}=\sum_{n\in Z}\hat{r}_{n}(t)e^{-i\lambda_{n}x}
\]
where
\[
\hat{r}_{n}(t)=\frac{1}{L}\int_{0}^{L}e^{\frac{i2\pi nx}{L}}r(t,x)dx=\frac{1}{L}\int_{0}^{L}e^{i\lambda_{n}x}r(t,x)dx
\]
Then from equation (\ref{hom_eq}), we get
\[
\hat{r}_{n}^{\prime\prime}+\alpha\hat{r}_{n}^{\prime}+\lambda_{n}^{2}\omega_{1}^{2}\hat{r}_{n}=0
\]
Solving this equation we find for $n\neq0$
\[
\hat{r}_{n}(t)=\hat{r}_{n}(0)(\tilde{a}(t\nu_{n})+\frac{\alpha\tilde{b}(t\nu_{n})}{2})+\hat{r}_{n}^{\prime}(0)\tilde{b}(t\nu_{n})
\]
For $n=0$ we have $\hat{r}_{0}(t)=0,$ as $\hat{r}_{0}^{\prime}(t)$
satisfies equation $\hat{r}_{0}^{\prime\prime}+\alpha\hat{r}_{0}^{\prime}=0$
and $\hat{r}_{0}(0)=\hat{r}_{0}^{\prime}(0)=0$ by (\ref{z_m}), (\ref{in_con}).

Then
\[
r(t,x)=\sum_{n:n\neq0}\hat{r}_{n}(t)e^{-i\lambda_{n}x}=
\]
\[
=\sum_{n\in Z:n\neq0}\hat{r}_{n}(0)e^{-i\lambda_{n}x}(\tilde{a}(t\nu_{n})+\frac{\alpha\tilde{b}(t\nu_{n})}{2})+\sum_{n\in Z:n\neq0}\hat{r}_{n}^{\prime}(0)e^{-i\lambda_{n}x}\tilde{b}(t\nu_{n})
\]
But by (\ref{in_con})
\[
\hat{r}_{n}(0)=\frac{1}{L}\int_{0}^{L}e^{\frac{i2\pi nx}{L}}X(x)dx,\quad\hat{r}_{n}^{\prime}(0)=\frac{1}{L}\int_{0}^{L}e^{\frac{i2\pi nx}{L}}V(x)dx
\]
The lemma is proved.

\begin{corollary}\label{cor}
For all $t>0$
\[
\int_{x}^{x+L}r(t,x^{\prime})dx^{\prime}=0
\]
\end{corollary}

We will consider also the nonhomogeneous wave equation with dissipation
\begin{equation}
G_{tt}(t,z)=\omega_{1}^{2}G_{zz}(t,z)-\alpha G_{t}(t,z)+f(t),z\in R\label{nonhom_eq}
\end{equation}
with initial conditions
\begin{equation}
G(0,z)=\int_{0}^{z}X(z^{\prime})dz^{\prime},\;G_{t}(0,z)=v+\int_{0}^{z}V(z^{\prime})dz^{\prime},v=\dot{x}_{0}(0)\label{b_c_n_hom}
\end{equation}
and the periodic boundary condition
\[
G(t,z+L)=G(t,z)+L
\]

\begin{lemma}\label{le2}There exists the only solution of (\ref{nonhom_eq}).
This solution has the following form
\begin{equation}
G(t,z)=G(t,0)+z+\int_{0}^{z}r(t,x)dx\label{g}
\end{equation}
where $r(t,x)$ is defined by (\ref{r}) and $G(t,0)$ is the solution
of ordinary differential equation 
\begin{equation}
G^{\prime\prime}(t,0)=-\alpha G^{\prime}(t,0)+\omega_{1}^{2}r_{x}(t,0)+f(t)\label{ode}
\end{equation}
with initial conditions $G(0,0)=0,$$G^{\prime}(0,0)=v.$
\end{lemma}

Note that solution of (\ref{ode}) is
\[
G^{\prime}(t,0)=e^{-\alpha t}v+\int_{0}^{t}e^{-\alpha(t-s)}(\omega_{1}^{2}r_{x}(s,0)+f(s))ds
\]
or
\begin{equation}
G(t,0)=\frac{1-e^{-\alpha t}}{\alpha}v+\frac{1}{\alpha}\int_{0}^{t}(1-e^{-\alpha(t-s)})f(s)ds+\frac{\omega_{1}^{2}}{\alpha}\int_{0}^{t}(1-e^{-\alpha(t-s)})r_{x}(s,0)ds\label{g0}
\end{equation}

\paragraph{Proof of lemma \ref{le2}}

We check by substitution that function $G(t,z),$ defined by (\ref{g})
satisfy equation (\ref{nonhom_eq}):
\[
G_{tt}=-\alpha G_{t}+\omega_{1}^{2}G_{zz}+f(t)
\]
By (\ref{g}) we have
\[
G_{t}(t,z)=G_{t}(t,0)+\int_{0}^{z}r_{t}(t,x)dx
\]
\[
G_{tt}(t,z)=G_{tt}(t,0)+\int_{0}^{z}r_{tt}(t,x)dx
\]
\[
G_{z}(t,z)=1+r(t,z),\quad G_{zz}(t,z)=r_{z}(t,z)=\int_{0}^{z}r_{xx}(t,x)dx+r_{z}(t,0)
\]
Substituting these expressions in (\ref{nonhom_eq}) we get identity
\begin{align*}
  G_{tt}(t,0)+\int_{0}^{z}r_{tt}(t,x)dx&=-\alpha G_{t}(t,0)-\alpha\int_{0}^{z}r_{t}(t,x)dx\\
  &\quad {} +\omega_{1}^{2}(\int_{0}^{z}r_{xx}(t,x)dx+r_{x}(t,0))+f(t).
\end{align*}
Indeed, by our condition $G(t,0)$ should satisfy equation (\ref{ode})
and 
\[
\int_{0}^{z}r_{tt}(t,x)dx=-\alpha\int_{0}^{z}r_{t}(t,x)dx+\omega_{1}^{2}\int_{0}^{z}r_{xx}(t,x)dx
\]
because of $r(t,x)$ is the solution of (\ref{hom_eq}).

Let us verify now the initial conditions:
\[
G(0,z)=G(0,0)+z+\int_{0}^{z}r(0,x)dx=z+\int_{0}^{z}(X(x)-1)dx=\int_{0}^{z}X(x)dx,
\]
\[
G_{t}(0,z)=G_{t}(0,0)+\int_{0}^{z}r_{t}(0,x)dx=v+\int_{0}^{z}V(x)dx,
\]
and the boundary condition:
\begin{align*}
  G(t,z+L)&=G(t,0)+z+L+\int_{0}^{z+L}r(0,x)dx\\
  &=G(t,0)+z+L+\int_{0}^{z}r(0,x)dx=G(t,z)+L
\end{align*}
as
\[
\int_{z}^{z+L}r(0,x)dx=0
\]
The lemma is proved.

\subsubsection{Convergence to continuum media}

\begin{theorem}\label{wave-1}
Let conditions of theorem \ref{absence_of_collisions} hold. Then

1) For any finite $T>0$ uniformly in $t\in[0,T]$ and in $z\in[0,L)$
\[
\lim_{N\to\infty}x_{[\frac{zN}{L}]}^{(N)}(t)=G(t,z),\:z\in[0,L)
\]

2) For any $T>0$ uniformly in $t\in[0,T]$ and in $x\in[0,L)$ there
exists the limit 
\begin{equation}
\lim_{N\to\infty}x_{k(x,N)}^{(N)}(t)=G(t,z(x))\label{w2-1}
\end{equation}
where $G(t,z)$ is the solution of (\ref{nonhom_eq}) which is given
by (\ref{g}) and (\ref{g0}).
\end{theorem}

\paragraph{Proof of theorem \ref{wave-1}}

Remind that
\[
r_{k}^{(N)}(t)=x_{k+1}^{(N)}(t)-x_{k}^{(N)}(t)-\frac{L}{N}=q_{k}(t)-\frac{L}{N}
\]
\[
\sum_{k=0}^{N-1}r_{k}^{(N)}(t)=0
\]
We begin with the following lemma

\begin{lemma} \label{differ} Uniformly in $t\in[0,T]$ for any finite
$T>0$
\[
\max_{t\in[0,T]}\max_{k=0,\ldots ,N-1}|r_{k}(t)-\frac{L}{N}r(t,\frac{kL}{N})|\leq\frac{C}{N^{3}},\quad N\to\infty
\]
where $r(t.x)$ is the solution of the equation (\ref{hom_eq}).
\end{lemma}

{\it Proof.} Consider the differences 
\[
\Delta_{k}^{(N)}(t)=r_{k}^{(N)}(t)-\frac{L}{N}r(t,\frac{kL}{N}),\quad k=0,\ldots ,N-1
\]
According to (\ref{q_equations}) we have the following system for
$r_{k},$$k=0,1,\ldots ,N-1$
\[
\ddot{r}_{k}^{(N)}(t)=\omega^{2}(r_{k+1}^{(N)}(t)-r_{k}^{(N)}(t)))-\omega^{2}(r_{k}^{(N)}(t)-r_{k-1}^{(N)}(t))-\alpha\dot{r}_{k}^{(N)}(t)
\]
with initial conditions
\[
r_{k}^{(N)}(0)=\frac{L}{N}X(\frac{kL}{N})+\xi_{k,N}-\frac{L}{N},\;\dot{r}_{k}^{(N)}(0)=\frac{L}{N}V(\frac{kL}{N})+\eta_{k,N}
\]
where 
\[
\xi_{k,N}=x_{k+1}(0)-x_{k}(0)-\frac{L}{N}X(\frac{kL}{N}),\quad \eta_{k,N}=\dot{x}_{k+1}(0)-\dot{x}_{k}(0)-\frac{L}{N}V(\frac{kL}{N})
\]
and, hence,
\[
\ddot{\Delta}_{k}^{(N)}(t)=\omega^{2}(r_{k+1}^{(N)}(t)-r_{k}^{(N)}(t))-\omega^{2}(r_{k}^{(N)}(t)-r_{k-1}^{(N)}(t))-\alpha\dot{r}_{k}^{(N)}(t)-\frac{L}{N}r_{tt}(t,\frac{kL}{N})
\]
As $r(t,x)$ satisfies the wave equation
\[
r_{tt}(t,\frac{kL}{N})=-\alpha r_{t}(t,\frac{kL}{N})+\omega_{1}^{2}r_{xx}(t,\frac{kL}{N})
\]
we get
\begin{align*}
  \ddot{\Delta}_{k}^{(N)}(t)&=\omega^{2}(r_{k+1}^{(N)}(t)-r_{k}^{(N)}(t)))-\omega^{2}(r_{k}^{(N)}(t)-r_{k-1}^{(N)}(t))-\alpha\dot{r}_{k}^{(N)}(t)\\
  &\quad {}-\frac{L}{N}(-\alpha r_{t}(t,\frac{kL}{N})+\omega_{1}^{2}r_{xx}(t,\frac{kL}{N}))=\\
&=\omega^{2}(r_{k+1}^{(N)}(t)-2r_{k}^{(N)}(t)+r_{k-1}^{(N)}(t))-\alpha\dot{\Delta}_{k}^{(N)}(t)-\frac{\omega_{1}^{2}L}{N}r_{xx}(t,\frac{kL}{N})
\end{align*}
and
\begin{align*}
  \ddot{\Delta}_{k}^{(N)}(t)&=\omega^{2}(r_{k+1}^{(N)}(t)-2r_{k}^{(N)}(t)+r_{k-1}^{(N)}(t))\\
  &\quad {} -\frac{\omega^{2}L}{N}(r(t,\frac{(k+1)L}{N})-2r(t,\frac{kL}{N})+r(t,\frac{(k-1)L}{N}))\\
                            &\quad {} +\frac{\omega^{2}L}{N}(r(t,\frac{(k+1)L}{N})-2r(t,\frac{kL}{N})+r(t,\frac{(k-1)L}{N}))\\
  &\quad {} -\alpha\dot{\Delta}_{k}^{(N)}(t)-\frac{\omega_{1}^{2}L}{N}r_{xx}(t,\frac{kL}{N})=\\
&=\omega^{2}(\Delta_{k+1}^{(N)}(t)-2\Delta_{k}^{(N)}(t)+\Delta_{k-1}^{(N)}(t))-\alpha\dot{\Delta}_{k}^{(N)}(t)+\\
                            &\quad {} +\omega_{0}^{2}LN(r(t,\frac{(k+1)L}{N})-2r(t,\frac{kL}{N})+r(t,\frac{(k-1)L}{N}))\\
  &\quad {} -\frac{\omega_{0}^{2}L^{3}}{N}r_{xx}(t,\frac{kL}{N})
\end{align*}
Define the remainder term as
\[
\delta_{k}^{(N)}(t)=r(t,\frac{(k+1)L}{N})-2r(t,\frac{kL}{N})+r(t,\frac{(k-1)L}{N})-\frac{L^{2}}{N^{2}}r_{xx}(t,\frac{kL}{N})
\]
Finally, we get the system of equations 
\begin{align*}
  &\ddot{\Delta}_{k}^{(N)}(t)=\omega^{2}(\Delta_{k+1}^{(N)}(t)-2\Delta_{k}^{(N)}(t)+\Delta_{k-1}^{(N)}(t))-\alpha\dot{\Delta}_{k}^{(N)}(t)+\omega_{0}^{2}LN\delta_{k}^{(N)}(t),\\
  &\hphantom{\ddot{\Delta}_{k}^{(N)}(t)=\omega^{2}(\Delta_{k+1}^{(N)}(t)-2\Delta_{k}^{(N)}(t)+\Delta_{k-1}^{(N)}(t))-\alpha\dot{\Delta}_{k}^{(N)}}
    k=0,1,\ldots ,N-1
\end{align*}
with initial conditions $\Delta_{k}^{(N)}(0)=\dot{\Delta}_{k}^{(N)}(0)=0.$
The solution
\[
\Delta_{k}^{(N)}(t)=\omega_{0}^{2}LN\int_{0}^{t}\tilde{b}((t-s)\nu_{k})\delta_{k}^{(N)}(s)ds
\]
where $\nu_{k}$ is defined by (\ref{nu}).

The remainder term can be estimated as follows
\[
|\delta_{k}^{(N)}(t)|\leq\frac{L^{4}}{12N^{4}}\max_{t\in[0,T]}\max_{x\in[0,L)}|r_{xxxx}(t,x)|=\frac{C_{0}}{N^{4}}
\]
It follows that
\[
|\Delta_{k}^{(N)}(t)|\leq\frac{C}{N^{3}}
\]
for some constant $C$ not depending on $N.$

We proceed to the proof of the theorem \ref{wave-1}.

1) We have the following equation for $\dot{x}_{0}(t)$
\[
\ddot{x}_{0}^{(N)}=-\alpha\dot{x}_{0}^{(N)}+\omega^{2}(r_{0}^{(N)}-r_{N-1}^{(N)})+f(t)
\]
with initial conditions $\dot{x}_{0}^{(N)}(0)=v.$ The solution is
\[
\dot{x}_{0}^{(N)}(t)=ve^{-\alpha t}+\int_{0}^{t}e^{-\alpha(t-s)}f(s)ds+\omega^{2}\int_{0}^{t}e^{-\alpha(t-s)}(r_{0}^{(N)}(s)-r_{N-1}^{(N)}(s))ds
\]
As 
\[
r_{0}^{(N)}(s)-r_{N-1}^{(N)}(s)=\frac{L}{N}(r(s,0)-r(s,\frac{L(N-1)}{N}))=
\]
\[
=\frac{L}{N}(r(s,0)-r(s,-\frac{L}{N}))=\frac{L^{2}}{N^{2}}r_{x}(s,0)+\frac{L}{N}\delta_{N}(s),
\]
where the remainder term can be estimated 
\[
|\delta_{N}(s)|\leq\frac{L^{2}}{2N^{2}}|r_{xx}(s,0)|,
\]
 we have
\[
\lim_{N\to\infty}\dot{x}_{0}^{(N)}(t)=ve^{-\alpha t}+\int_{0}^{t}e^{-\alpha(t-s)}f(s)ds+\omega_{0}^{2}L^{2}\int_{0}^{t}e^{-\alpha(t-s)}r_{x}(s,0)ds
\]
and
\begin{align*}
  \lim_{N\to\infty}x_{0}^{(N)}(t)&=v\frac{1-e^{-\alpha t}}{\alpha}+\int_{0}^{t}(1-e^{-\alpha(t-s)})f(s)ds\\
  &\quad {} +\frac{\omega_{0}^{2}L^{2}}{\alpha}\int_{0}^{t}(1-e^{-\alpha(t-s)})r_{x}(s,0)ds
\end{align*}
By (\ref{g0})
\[
\lim_{N\to\infty}x_{0}^{(N)}(t)=G(t,0)
\]
Further on, we have
\[
x_{[\frac{zN}{L}]}^{(N)}(t)=x_{0}^{(N)}(t)+[\frac{zN}{L}]\frac{L}{N}+\sum_{j=0}^{[\frac{zN}{L}]-1}r_{j}^{(N)}(t)=
\]
\[
=x_{0}^{(N)}(t)+z+\sum_{j=0}^{[\frac{zN}{L}]-1}\frac{L}{N}r(t,\frac{jL}{N})+\delta^{(N)}(t,z)
\]
where 
\[
\delta^{(N)}(t,z)=\sum_{j=0}^{[\frac{zN}{L}]-1}(r_{j}^{(N)}(t)-\frac{L}{N}r(t,\frac{jL}{N}))=\sum_{j=0}^{[\frac{zN}{L}]-1}\Delta_{j}^{(N)}(t)
\]
One can conclude from the proof of lemma \ref{differ} 
\[
|\delta^{(N)}(t,z)|\leq\frac{zC}{LN^{2}}
\]

Hence,
\begin{align*}
  \lim_{N\to\infty}x_{[\frac{zN}{L}]}^{(N)}(t)&=G(t,z)\\
  &=\lim_{N\to\infty}x_{0}^{(N)}(t)+z+\int_{0}^{z}r(t,z^{\prime})dz^{\prime}=G(t,0)+z+\int_{0}^{z}r(t,z^{\prime})dz^{\prime}
\end{align*}
By (\ref{g})
\[
\lim_{N\to\infty}x_{[\frac{zN}{L}]}^{(N)}(t)=G(t,z)
\]

2) Let us prove that for some constant $d>0$
\[
|\frac{Lk(x,N)}{N}-z(x)|\leq\frac{d}{N}
\]
uniformly in $x\in[0,L).$ Denote 
\[
h(z)=\int_{0}^{z}X(x^{\prime})dx^{\prime}
\]
Then we have $h(z(x))=x$. On the other side, the integral can be
calculated as follows
\[
h(\frac{Lk(x,N)}{N})=\frac{L}{N}\sum_{i=0}^{k(x,N)}X(\frac{iL}{N})+s_{N}(x)=x_{k(x,N)+1}^{(N)}(0)-\sum_{i=0}^{k(x,N)}\xi_{i,N}+s_{N}(x)
\]
where
\[
s_{N}(x)=h(\frac{Lk(x,N)}{N})-\frac{L}{N}\sum_{i=0}^{k(x,N)}X(\frac{iL}{N})
\]
\[
\xi_{i,N}=x_{i+1}^{(N)}(0)-x_{i}^{(N)}(0)-\frac{L}{N}X(\frac{iL}{N}).
\]
 By (\ref{regularity_cond}) the remainder terms enjoys the following
estimate:
\[
|\sum_{l=0}^{k(x,N)}\xi_{i,N}|=O(N^{-1})
\]
\[
|s_{N}(x)|\leq\frac{L}{N}\max_{y\in[0,L)}|X^{\prime}(y)|=\frac{d^{\prime}}{N}.
\]
So
\[
|h(\frac{Lk(x,N)}{N})-x_{k(x,N)+1}^{(N)}(0)|\leq\frac{d_{1}}{N}
\]
for some constant $d_{1}.$ 

By (\ref{k_x_N}) we have:
\begin{align*}
  x-\frac{d_{1}}{N}&\leq h(\frac{Lk(x,N)}{N})\\
  &<x+\frac{L}{N}X(\frac{Lk(x,N)}{N})+\frac{d_{1}}{N}<x+\frac{d_{2}}{N},d_{2}=d_{1}+\max_{y\in[0,L)}|X(y)|
\end{align*}
It follows that
\[
|h(\frac{Lk(x,N)}{N})-h(z(x))|\leq\frac{d_{2}}{N}
\]
For some $\theta\in[0,L)$ 
\[
h(\frac{Lk(x,N)}{N})-h(z(x))=(\frac{Lk(x,N)}{N}-z(x))h^{\prime}(\theta)
\]
where $h^{\prime}(\theta)=X(\theta).$ This gives
\begin{equation}
|\frac{Lk(x,N)}{N}-z(x)|\leq\frac{d_{2}}{N\min_{\theta\in[0,L)}|X(\theta)|}=\frac{d}{N}\label{kzd}
\end{equation}
From the proved inequality it follows that
\[
|k(x,N)-[\frac{z(x)N}{L}]|\leq\frac{d}{L}+1=d^{\prime}
\]
By theorem \ref{absence_of_collisions}
\begin{equation}
|x_{k(x,N)}^{(N)}(t)-x_{[\frac{z(x)N}{L}]}^{(N)}(t)|\leq\frac{d^{\prime}L(1+\delta)}{N}\label{diff}
\end{equation}
Taking the limit in the last inequality and using item 1) of the theorem
we get assertion 2).

Now we can finish the proof of theorem \ref{wave}. The first item
of this theorem follows from the second assertion of theorem \ref{wave-1}.

Let us prove that $Y(t,x)=g(t,z(x))$ is strictly increasing over
$x\in R$. Let $z_{1}<z_{2}$. From evident equality 
\[
x_{[\frac{z_{2}N}{L}]}^{(N)}(t)-x_{[\frac{z_{1}N}{L}]}^{(N)}(t)=\sum_{k=[\frac{z_{1}N}{L}]}^{[\frac{z_{2}N}{L}]-1}x_{k+1}^{(N)}(t)-x_{k}^{(N)}(t)
\]
 and from theorem \ref{absence_of_collisions} we have
\[
\frac{L(1-\delta)}{N}([\frac{z_{2}N}{L}]-[\frac{z_{1}N}{L}])\leq x_{[\frac{z_{2}N}{L}]}^{(N)}(t)-x_{[\frac{z_{1}N}{L}]}^{(N)}(t)\leq\frac{L(1+\delta)}{N}([\frac{z_{2}N}{L}]-[\frac{z_{1}N}{L}]).
\]
Taking the limit here and using theorem \ref{wave-1} we get for any
$t>0$ and any $z_{1}<z_{2}$
\begin{equation}
L(1-\delta)(z_{2}-z_{1})\leq G(t,z_{2})-G(t,z_{1})\leq L(1+\delta)(z_{2}-z_{1})\label{mon}
\end{equation}
So function $Y(t,x)=G(t,z(x)),$ $x\in R$ is strictly increasing
and differentiable with respect to $x$
\[
G_{x}(t,z(x))=z^{\prime}(x)(1+r(t,z(x))
\]
 Thus, $Y(t,x)$ is a diffeomorphism of $R$. 

\subsection{Convergence to Euler equation}

\subsubsection{Proof of lemma \ref{density}}

Note that there always exists $m=m(t)$ such that
\[
x_{k}^{(N)}(t)\in[0,L),k=m,m+1,\ldots ,m+N-1
\]
Then
\[
F^{(N)}(t,y)=\frac{1}{N}\sharp\{k=m,m+1,\ldots ,m+N-1:x_{k}^{(N)}(t)\leq y\},\quad y\in[0,L)
\]
For given $y$ and $t$ one can define the number $k(y,N,t)$ such
that 
\begin{equation}
x_{k(y,N,t)}^{(N)}(t)\leq y<x_{k(y,N,t)+1}^{(N)}(t),\quad y\in[0,L)\label{k_y_N_t}
\end{equation}
where $m\leq k(y,N,t)<m+N.$ So
\[
F^{(N)}(t,y)=\frac{k(y,N,t)-m(t)}{N}
\]
We use the evident inequality
\begin{align*}
  |\hat{x}_{k(x(t,y),N)}^{(N)}(t)-x_{k(y,N,t)}^{(N)}(t)|&\leq|\hat{x}_{k(x(t,y),N)}^{(N)}(t)-y(t,x(t,y))|\\
  &\quad {} +|x_{k(y,N,t)}^{(N)}(t)-y(t,x(t,y))|
\end{align*}
By assertions 1), 2) of theorem \ref{wave-1}, and by (\ref{diff}),
one can conclude, that 
\[
|\hat{x}_{k(x(t,y),N)}^{(N)}(t)-y(t,x(t,y)))|\leq\frac{d_{1}}{N}
\]
But $\hat{x}_{k(x(t,y),N)}^{(N)}(t)=x_{k(x(t,y),N)+m(t)}^{(N)}(t)$
\[
|x_{k(x(t,y),N)+m(t)}^{(N)}(t)-y(t,x(t,y)))|\leq\frac{d_{1}}{N},
\]
for some constant $d_{1}>0$ not depending on $N$ and $y\in[0,L)$.
By (\ref{k_y_N_t}) and theorem \ref{absence_of_collisions} we have 

\[
|x_{k(y,N,t)}^{(N)}(t)-y|\leq|x_{k(y,N,t)}^{(N)}(t)-x_{k(y,N,t)+1}^{(N)}(t)|\leq\frac{d_{2}}{N},
\]
for some constant $d_{2}>0$ not depending on $N,y$. Then
\[
|x_{k(x(t,y),N)+m(t)}^{(N)}(t)-x_{k(y,N,t)}^{(N)}(t)|\leq\frac{d_{1}+d_{2}}{N}
\]
From this inequality and theorem \ref{absence_of_collisions} we have
\begin{equation}
|k(x(t,y),N)+m(t)-k(y,N,t)|=|k(x(t,y),N)-(k(y,N,t)-m(t))|\leq d^{\prime}\label{kkd}
\end{equation}
for $d^{\prime}=d_{1}+d_{2}>0$, not depending on $N,y$. We can conclude
that 
\[
\lim_{N\rightarrow\infty}F^{(N)}(t,y)=\lim_{N\rightarrow\infty}\frac{k(y,N,t)-m(t)}{N}=\lim_{N\rightarrow\infty}\frac{k(x(t,y),N)}{N}=\frac{z(x(t,y))}{L}
\]
where the latter equality follows from (\ref{kzd}). The lemma is
proved.

\subsubsection{Proof of theorem \ref{euler}}

Let us prove (\ref{cons_law}). By (\ref{distr_fun}) and (\ref{den})
we have
\begin{equation}
\rho(t,y)=L^{-1}z^{\prime}(x(t,y))x_{y}(t,y).\label{density_formula}
\end{equation}
On the other side, differentiation in $y$ of the equality $y(t,x(t,y))=y$
gives
\[
x_{y}(t,y)=\frac{1}{y_{x}(t,x(t,y))}.
\]
Hence,
\[
y_{x}(t,x(t,y))=\frac{z^{\prime}(x(t,y))}{L\rho(t,y)}
\]
By (\ref{den}) and lemma \ref{density} we have
\[
\frac{\partial\rho(t,y)}{\partial t}=L^{-1}\frac{d}{dy}\frac{dz(x(t,y))}{dt}=L^{-1}\frac{d}{dy}(z^{\prime}(x(t,y))x_{t}(t,y))
\]
Differentiation in $t$ of the equality $y(t,x(t,y))=y$ gives
\[
\frac{\partial y(t,x(t,y))}{\partial t}+y_{x}(t,x(t,y))x_{t}(t,y)=0
\]
So
\[
x_{t}(t,y)=-\frac{\frac{\partial y(t,x(t,y))}{\partial t}}{y_{x}(t,x(t,y))}=-\frac{u(t,y)}{y_{x}(t,x(t,y))}=-\frac{u(t,y)L\rho(t,y)}{z^{\prime}(x(t,y))}
\]
and
\[
\frac{\partial\rho(t,y)}{\partial t}=-\frac{d}{dy}(u(t,y)\rho(t,y))
\]

To prove (\ref{Euler}) note that
\[
\frac{\partial u(t,y)}{\partial t}+u(t,y)\frac{\partial u(t,y)}{\partial y}=\frac{\partial u(t,y(t,x))}{\partial t}=\frac{\partial^{2}y(t,x)}{\partial t^{2}}
\]
On the other side`
\[
\frac{\partial u(t,y(t,x))}{\partial t}=\frac{\partial^{2}y(t,x)}{\partial t^{2}}
\]
By theorem \ref{wave-1} we have
\begin{equation}
y(t,x)=G(t,z(x)) \mod L \label{y_g}
\end{equation}
\[
G(t,z(x+L))=G(t,z(x)+L)=G(t,z(x))+L
\]
\[
G(t,z(L))=G(t,L)=G(t,0)+L
\]
where $G(t,z)$ satisfies the equation 
\[
G_{tt}(t,z)=-\alpha G_{t}(t,z)+\omega_{1}^{2}G_{zz}(t,z)+f(t)
\]
where $\omega_{1}=\omega_{0}L.$ Using these formulas we find
\begin{align}
  \frac{\partial u(t,y)}{\partial t}+u(t,y)\frac{\partial u(t,y)}{\partial y}&=\frac{\partial^{2}y(t,x)}{\partial t^{2}}=G_{tt}(t,z(x))\notag \\
  &=-\alpha G_{t}(t,z(x))+\omega_{1}^{2}G_{zz}(t,z(x))+f(t)\label{u_t}
\end{align}
 Further on, using formula (\ref{y_g}) let us calculate derivatives
\begin{equation}
y_{t}(t,x)=\frac{\partial G(t,z(x))}{\partial t}=G_{t}(t,z(x))\label{y_t}
\end{equation}
\[
y_{x}(t,x)=\frac{\partial G(t,z(x))}{\partial x}=z\prime(x)\frac{\partial G(t,z(x))}{\partial z}=z\prime(x)G_{z}(t,z(x))
\]
\begin{align*}
  y_{xx}(t,x)&=[z\prime(x)]^{2}\frac{\partial^{2}G(t,z(x))}{\partial z^{2}}+z\prime\prime(x)\frac{\partial G(t,z(x))}{\partial z}\\
  &=[z\prime(x)]^{2}G_{zz}(t,z(x))+z\prime\prime(x)G_{z}(t,z(x)).
\end{align*}
It follows that
\[
G_{zz}(t,z(x))=\frac{y_{xx}(t,x)-z\prime\prime(x)G_{z}(t,z(x))}{[z\prime(x)]^{2}}=\frac{y_{xx}(t,x)-\frac{z\prime\prime(x)}{z\prime(x)}y_{x}(t,x)}{[z\prime(x)]^{2}}
\]
So we get 
\[
\frac{\partial u(t,y(t,x))}{\partial t}+u(t,y)\frac{\partial u(t,y(t,x))}{\partial y}=-\alpha G_{t}(t,z(x))+\omega_{1}^{2}G_{zz}(t,z(x))+f(t)=
\]
\[
=-\alpha y_{t}(t,x)+\omega_{1}^{2}\frac{y_{xx}(t,x)-\frac{z\prime\prime(x)}{z\prime(x)}y_{x}(t,x)}{[z\prime(x)]^{2}}+f(t)
\]
Putting in this equation $x=x(t,y)$ and defining function
\begin{equation}
R(t,y)=G_{zz}(t,z(x(t,y)))=\frac{y_{xx}(t,x(t,y))-\frac{z\prime\prime(x(t,y))}{z\prime(x(t,y))}y_{x}(t,x(t,y))}{[z\prime(x(t,y))]^{2}}\label{R}
\end{equation}
we get 
\begin{align*}
  \frac{\partial u(t,y)}{\partial t}+u(t,y)\frac{\partial u(t,y)}{\partial y}&=-\alpha y_{t}(t,x(t,y))+\omega_{1}^{2}R(t,y)+f\\
  &=-\alpha u(t,y)+\omega_{1}^{2}R(t,y)+f(t)
\end{align*}
Differentiating in $y$
\[
y_{xx}(t,x(t,y))=\frac{1}{x_{y}(t,y)}\frac{d}{dy}\frac{z^{\prime}(x(t,y))}{\rho(t,y)}=\frac{z^{\prime}(x(t,y))}{\rho(t,y)}\frac{d}{dy}\frac{z^{\prime}(x(t,y))}{\rho(t,y)}=
\]
\[
=\frac{z^{\prime}(x(t,y))}{\rho(t,y)}\left(\frac{z^{\prime\prime}(x(t,y))x_{y}(t,y)}{\rho(t,y)}-\frac{z^{\prime}(x(t,y))\rho_{y}(t,y)}{\rho^{2}(t,y)}\right)=
\]
\[
=\frac{z^{\prime}(x(t,y))}{\rho(t,y)}\left(\frac{z^{\prime\prime}(x(t,y))}{z^{\prime}(x(t,y))}-\frac{z^{\prime}(x(t,y))\rho_{y}(t,y)}{\rho^{2}(t,y)}\right)=
\]
\[
=\frac{1}{\rho(t,y)}\left(z^{\prime\prime}(x(t,y))-\frac{(z^{\prime}(x(t,y)))^{2}\rho_{y}(t,y)}{\rho^{2}(t,y)}\right)
\]
So the function $R(t,y)$ can be written as
\[
R(t,y)=\frac{y_{xx}(t,x(t,y))-\frac{z\prime\prime(x(t,y))}{z\prime(x(t,y))}y_{x}(t,x(t,y))}{[z\prime(x(t,y))]^{2}}=
\]
\[
=\frac{\frac{1}{\rho(t,y)}\left(z^{\prime\prime}(x(t,y))-\frac{(z^{\prime}(x(t,y)))^{2}\rho_{y}(t,y)}{\rho^{2}(t,y)}\right)-\frac{z\prime\prime(x(t,y))}{z\prime(x(t,y))}\frac{z^{\prime}(x(t,y))}{\rho(t,y)}}{[z\prime(x(t,y))]^{2}}=
\]
\[
=\frac{1}{\rho(t,y)}\frac{z^{\prime\prime}(x(t,y))-\frac{(z^{\prime}(x(t,y)))^{2}\rho_{y}(t,y)}{\rho^{2}(t,y)}-z\prime\prime(x(t,y))}{[z\prime(x(t,y))]^{2}}=-\frac{\rho_{y}(t,y)}{\rho^{3}(t,y)}=
\]
\[
=\frac{1}{\rho(t,y)}\frac{d}{dy}\frac{1}{\rho(t,y)}
\]
By (\ref{pressure}) 
\begin{equation}
\omega_{1}^{2}R(t,y)=\frac{1}{\rho(t,y)}\frac{d}{dy}\frac{\omega_{1}^{2}}{\rho(t,y)}=-\frac{p_{y}(t,y)}{\rho(t,y)}\label{R_1}
\end{equation}
Finally, we come to the equation
\begin{align*}
  \frac{\partial u(t,y)}{\partial t}+u(t,y)\frac{\partial u(t,y)}{\partial y}&=-\alpha u(t,y)-\frac{\omega_{1}^{2}\rho_{y}(t,y)}{\rho^{3}(t,y)}+f(t)\\
  &=-\alpha u(t,y)-\frac{p_{y}(t,y)}{\rho(t,y)}+f(t)
\end{align*}
The theorem is proved.

\subsubsection{Proof of theorem \ref{Euler_Lagrange}}

Let us prove (\ref{fe}). By (\ref{density_formula}) we have 
\[
\rho(t,y)=L^{-1}z^{\prime}(x(t,y))x_{y}(t,y).
\]
On the other side, differentiation in $y$ of the equality $y(t,x(t,y))=y$
gives
\[
x_{y}(t,y)=\frac{1}{y_{x}(t,x(t,y))}.
\]
Hence,
\[
\rho(t,y)=\frac{z^{\prime}(x(t,y))}{Ly_{x}(t,x(t,y))}
\]
and by (\ref{first})
\[
\hat{\rho}(t,z)=L\rho(t,y(t,x(z)))=\frac{Lz^{\prime}(x(t,y(t,x(z))))}{Ly_{x}(t,x(t,y(t,x(z))))}=\frac{z^{\prime}(x(z))}{y_{x}(t,x(z))}=
\]
\[
=\frac{z^{\prime}(x(z))x^{\prime}(z)}{y_{x}(t,x(z))x^{\prime}(z)}=\frac{1}{G_{z}(t,z)}
\]
as $G_{z}(t,z)=y_{x}(t,x(z))x^{\prime}(z)$ and $z^{\prime}(x(z))x^{\prime}(z)=1.$
So
\begin{equation}
\frac{1}{\hat{\rho}(t,z)}=G_{z}(t,z)\label{Gz-1}
\end{equation}
\[
\frac{\partial}{\partial t}\left(\frac{1}{\hat{\rho}(t,z)}\right)=\frac{\partial^{2}G(t,z)}{\partial t\partial z}
\]
Further on, by (\ref{second})
\[
\frac{\partial\hat{u}(t,z)}{\partial z}=\frac{\partial^{2}y(t,x(z))}{\partial z\partial t}=\frac{\partial^{2}G(t,z)}{\partial z\partial t}=\frac{\partial^{2}G(t,z)}{\partial t\partial z}
\]
Thus, we come to the equation
\[
\frac{\partial}{\partial t}\left(\frac{1}{\hat{\rho}(t,z)}\right)-\frac{\partial\hat{u}(t,z)}{\partial z}=0
\]

To prove (\ref{se}) note that by (\ref{second})
\[
\frac{\partial\hat{u}(t,z)}{\partial t}=\frac{\partial^{2}y(t,x(z))}{\partial t^{2}}=\frac{\partial^{2}G(t,z)}{\partial t^{2}}
\]
By theorem \ref{wave-1} we have
\[
y(t,x(z))=G(t,z)\,(\mod L)
\]
where $G(t,z)$ satisfies the wave equation 
\[
G_{tt}(t,z)=-\alpha G_{t}(t,z)+\omega_{1}^{2}G_{zz}(t,z)+f
\]
Using these formulas we find
\[
\frac{\partial\hat{u}(t,z)}{\partial t}=-\alpha G_{t}(t,z)+\omega_{1}^{2}G_{zz}(t,z)+f
\]
By (\ref{second}), (\ref{Gz-1})
\[
\hat{u}(t,z)=G_{t}(t,z)
\]
\[
\omega_{1}^{2}G_{zz}(t,z)=\frac{\partial}{\partial z}\left(\frac{\omega_{1}^{2}}{\hat{\rho}(t,z)}\right)=-\frac{\partial}{\partial z}\left(-\frac{\omega_{1}^{2}}{\hat{\rho}(t,z)}\right)=-\frac{\partial\hat{p}(t,z)}{\partial z}
\]
where 
\[
\hat{p}(t,z)=-\frac{\omega_{1}^{2}}{\hat{\rho}(t,z)}+C
\]
So, we come to the second equation
\[
\frac{\partial\hat{u}(t,z)}{\partial t}+\alpha\hat{u}(t,z)-f=-\frac{\partial\hat{p}(t,z)}{\partial z}
\]
The theorem is proved.

\subsubsection{Proof of theorem \ref{force}}

By (\ref{r_n}) we have
\[
R^{(N)}(t,y)=\omega^{2}(r_{k(y,N,t)}^{(N)}(t)-r_{k(y,N,t)-1}^{(N)}(t))
\]
By lemma \ref{differ}
\[
R^{(N)}(t,y)=\omega_{0}^{2}N^{2}\frac{L}{N}(r(t,\frac{k(y,N,t)L}{N})-r(t,\frac{(k(y,N,t)-1)L}{N})+O(N^{-1})=
\]
\[
=\omega_{0}^{2}L^{2}r_{x}(t,\frac{k(y,N,t)L}{N})+O(N^{-1})=\omega_{1}^{2}r_{x}(t,\frac{k(y,N,t)L}{N})+O(N^{-1})
\]

Using inequalities (\ref{kzd}) and (\ref{kkd}) we have the following
estimate:
\[
|\frac{k(y,N,t)L}{N}-z(x(t,y))|\leq\frac{c}{N}
\]
for some constant $c$, not depending on $N$. Then we can conclude
that
\[
\lim_{N\rightarrow\infty}R^{(N)}(t,y)=\omega_{1}^{2}r_{x}(t,z(x(t,y)))=\omega_{1}^{2}G_{zz}(t,z(x(t,y)))=\omega_{1}^{2}R(t,y)
\]
Using (\ref{R}) and (\ref{R_1}) we get the assertion of the theorem.

\subsubsection{Proof of theorem \ref{wave_eq}}

The item 2) follows from theorem \ref{wave-1}. To prove the item
1), substitute $G(t,z)=Y(t,x(z))$ into equation
\[
G_{tt}(t,z)=\omega_{1}^{2}G_{zz}(t,z)-\alpha G_{t}(t,z)+f(t)
\]
The function $x(z)$ is inverse to $z(x)$, which is defined by (\ref{Z_x}).
So 
\[
x(z)=\int_{0}^{z}X(x^{\prime})dx^{\prime}
\]
 Calculating derivatives
\[
G_{tt}(t,z)=Y_{tt}(t,x(z))
\]
\[
G_{zz}(t,z)=Y_{xx}(t,x(z))X^{2}(z)+Y_{x}(t,x(z))X^{\prime}(z)
\]
we come to the desired equation
\[
Y_{tt}(t,x(z))=\omega_{1}^{2}\left(Y_{xx}(t,x(z))X^{2}(z)+Y_{x}(t,x(z))X^{\prime}(z)\right)-\alpha Y_{t}(t,x(z))+f(t)
\]

\subsubsection{Proof of theorem \ref{exp_dyn}}

Consider the nonhomogeneous wave equation
\[
g_{tt}(t,z)=\omega_{1}^{2}g_{zz}(t,z)-\alpha g_{t}(t,z)+f(t)
\]
with initial conditions
\[
g(0,z)=\phi(z)=\int_{0}^{z}X(u)du,\;g_{t}(0,z)=\psi(z)=v+\int_{0}^{z}V(u)du,v=\dot{x}_{0}(0)
\]

The substitution $g(t,z)=e^{-\frac{\alpha}{2}t}w(t,z)$ leads to the
 equation
\[
w_{tt}(t,z)=\omega_{1}^{2}w_{zz}(t,z)+\frac{\alpha^{2}}{4}w(t,z)+e^{\frac{\alpha}{2}t}f(t)
\]
with initial conditions 
\[
w(0,z)=\phi(z),\;w_{t}(0,z)=\psi(z)+\frac{\alpha}{2}\phi(z)
\]
The solution of this equation has the form (see \cite{Pol}, p.\thinspace 569)
\[
w(t,z)=\frac{1}{2}(\phi(z+\omega_{1}t)+\phi(z-\omega_{1}t))+
\]
\[
+\frac{\alpha t}{4\omega_{1}}\int_{z-\omega_{1}t}^{z+\omega_{1}t}\frac{I_{1}(\frac{\alpha}{2}\sqrt{t^{2}-(z-\xi)^{2}/\omega_{1}^{2}})}{\sqrt{t^{2}-(z-\xi)^{2}/\omega_{1}^{2}}}\phi(\xi)d\xi+
\]
\[
+\frac{1}{2\omega_{1}}\int_{z-\omega_{1}t}^{z+\omega_{1}t}I_{0}(\frac{\alpha}{2}\sqrt{t^{2}-(z-\xi)^{2}/\omega_{1}^{2}})(\psi(\xi)+\frac{\alpha}{2}\phi(\xi))d\xi+
\]
\[
+\frac{1}{2\omega_{1}}\int_{0}^{t}\int_{z-\omega_{1}(t-\tau)}^{z+\omega_{1}(t-\tau)}I_{0}(\frac{\alpha}{2}\sqrt{(t-\tau)^{2}-(z-\xi)^{2}/\omega_{1}^{2}})e^{\frac{\alpha}{2}\tau}f(\tau)d\xi d\tau
\]
Thus, 
\[
g(t,z)=e^{-\frac{\alpha}{2}t}w(t,z)=\frac{e^{-\frac{\alpha}{2}t}}{2}(\phi(z+\omega_{1}t)+\phi(z-\omega_{1}t))+
\]
\[
+\frac{\alpha e^{-\frac{\alpha}{2}t}}{4\omega_{1}}\int_{z-\omega_{1}t}^{z+\omega_{1}t}\Bigl(t\frac{I_{1}(\frac{\alpha}{2}\sqrt{t^{2}-(z-\xi)^{2}/\omega_{1}^{2}})}{\sqrt{t^{2}-(z-\xi)^{2}/\omega_{1}^{2}}}+I_{0}(\frac{\alpha}{2}\sqrt{t^{2}-(z-\xi)^{2}/\omega_{1}^{2}})\Bigr)\phi(\xi)d\xi+
\]
\[
+\frac{e^{-\frac{\alpha}{2}t}}{2\omega_{1}}\int_{z-\omega_{1}t}^{z+\omega_{1}t}I_{0}(\frac{\alpha}{2}\sqrt{t^{2}-(z-\xi)^{2}/\omega_{1}^{2}})\psi(\xi)d\xi+
\]
\[
+\frac{1}{2\omega_{1}}\int_{0}^{t}e^{-\frac{\alpha}{2}(t-\tau)}f(\tau)d\tau\int_{z-\omega_{1}(t-\tau)}^{z+\omega_{1}(t-\tau)}I_{0}(\frac{\alpha}{2}\sqrt{(t-\tau)^{2}-(z-\xi)^{2}/\omega_{1}^{2}})d\xi
\]
Here $I_{0}(x),I_{1}(x)$ are modified Bessel functions:
\[
I_{0}(x)=J_{0}(ix)=\sum_{m=0}^{\infty}\frac{1}{m!\Gamma(m+1)}\left(\frac{x}{2}\right)^{2m}
\]
\[
I_{1}(x)=i^{-1}J_{1}(ix)=\sum_{m=0}^{\infty}\frac{1}{m!\Gamma(m+2)}\left(\frac{x}{2}\right)^{2m+1}
\]

\end{document}